\documentclass[prd,aps,a4paper,nofootinbib,twocolumn]{revtex4}  

\newif\ifusesec
\usesectrue  
   
\usepackage{graphicx} 
\usepackage{mathrsfs}
\usepackage{amsmath,amsfonts,amssymb}
\usepackage{multirow}

\newcommand{\beq}{\begin{equation}}
\newcommand{\eeq}{\end{equation}}

\begin{document}

\title{G\"odel spacetime: elliptic-like geodesics and gyroscope precession}

\author{Donato \surname{Bini}$^{1,2}$, Andrea \surname{Geralico}$^1$, Robert T. Jantzen$^3$, Wolfango \surname{Plastino}$^{2,4}$}

\affiliation{$^1$Istituto per le Applicazioni del Calcolo ``M. Picone'', CNR, I-00185 Rome, Italy\ }
\affiliation{$^2$INFN, Sezione di Roma Tre, I-00146 Rome, Italy\ }
\affiliation{$^3$Department of Mathematics and Statistics, Villanova University, Villanova, PA 19085, USA}
\affiliation{$^4$Roma Tre University, Department of Mathematics and Physics, I-00146 Rome, Italy\ }

\begin{abstract}
We study elliptic-like geodesic motion on hyperplanes orthogonal to the cylindrical symmetry axes of the G\"odel spacetime by using an eccentricity-semi-latus rectum parametrization which is familiar from the Newtonian description of a two-body system.  We compute several quantities which summarize the main features of the motion, namely the coordinate time and proper time periods of the radial motion, the frequency of the azimuthal motion, the full variation of the azimuthal angle over a period, etc. Exact as well as approximate (i.e., Taylor-expanded in the limit of small eccentricity) analytic expressions of all these quantities are obtained.
Finally, we consider their application to the gyroscope precession frequency along these orbits, generalizing the existing results for the circular case.
\end{abstract}
 
\date{\today}

\maketitle

\section{Introduction}

The G\"odel dust cosmological solution \cite{Godel:1949ga} of the Einstein equations dates back to 1949, and ever since it has captured people's imagination as a globally rotating universe model. Besides its main geometric characterization (briefly summarized below), this solution has an important physical meaning in terms of the associated gravitational field: the dust source particles are at rest with respect to the coordinates (in the standard choice of coordinates with which the metric is commonly expressed), but form a family of twisting world lines, in such a way that the  cylindrical symmetry of the spacetime is preserved about every point.

This spacetime has been widely studied \cite{Hawell}. Its spectral classification, geodesics and causal structure are all known \cite{chandra,Novello:1982nc,Iyer:1993qa,Rindler:1990}. A nontrivial consequence of this last analysis is the existence of closed timelike lines (CTCs), which led to mostly discarding the G\"odel solution from the set of ``good solutions," leaving it primarily to be of mathematical rather than physical interest.

However, the opposite point of view can also be adopted. Namely,  if the features of a strong gravitational field are to be systematically explored, then all exact solutions, not only those more familiar ones like black holes or gravitational waves which have proven to be very useful to make our intuition more robust, can play a role, including spacetimes like G\"odel's with pathological regions where time travel is allowed, especially because it might be difficult to disentangle these pathologies from the strong field features with a conjecture (like the Penrose cosmic censorship conjecture).

Timelike circular orbits have been thoroughly studied and are rather trivial, which is the case for all similar G\"odel-like solutions obtained in general relativity as well as in other extended gravitational theories \cite{Calvao:1990yv,Gleiser:2005nw}.
The aim of the present article is to analyze those timelike geodesics~\cite{Novello:1982nc} which are not circular, but which may reduce to circular orbits in the limit of vanishing (suitably defined) eccentricity, within a coordinate system standardly adapted to the cylindrical symmetry of the spacetime.
We have 1) derived conditions which allow the existence of such orbits on a ``planar" 2-surface orthogonal to the cylindrical symmetry axes (in reality an intrinsically curved surface, but extrinsically flat within the time coordinate hypersurfaces), 2) determined the associated minimum and maximum radii, i.e., the ``periastron" and ``apastron," and 3) found a novel extension of the classical semi-latus rectum ($p$) and eccentricity ($e$) parametrization of these orbits which allows for a complete analytical treatment as well as an estimate of the ``distance from the circular configuration" using a Taylor series in the eccentricity. By doing so we have 4) mapped the standard timelike geodesic parameter space described by the conserved energy $(E)$ and angular momentum ($L$) onto $p$ and $e$. 
Furthermore, along those closed eccentric orbits which encircle the cylindrical symmetry axis, we have 5) introduced two physical frames: a Frenet-Serret frame and a parallel transported frame (recently obtained in Appendix B of Ref.~\cite{Bini:2018zfu}), allowing us to discuss test gyroscope precession (a novel result of the  present paper)  together with its limit of the known circular orbit case~\cite{Iyer:1993qa,Rindler:1990,Wilkins:1992}. 

The paper is organized as follows: Sec.~II recalls the main geometrical properties of the G\"odel spacetime, including its geodesics and their classification in the parameter space \cite{Novello:1982nc}. Sec.~III analyzes the timelike geodesics on the planar surfaces orthogonal to the cylindrical symmetry axes, in order to study gyroscopic precession in Sec.~IV, completing previous results \cite{Bini:2018zfu}. The signature of the metric is $-+++$ and the units are chosen so that $c=1=G$.

\section{G\"odel spacetime: geometrical properties}

The G\"odel spacetime \cite{Godel:1949ga,Hawell} is a Petrov type D stationary axisymmetric solution of the nonvacuum Einstein equations with a cosmological constant whose metric when expressed in cylindrical coordinates $(t,r,\phi,z)$ reads\footnote{
We use the form of the metric given by Ref.~\cite{Hawell}. Appendix A shows how this follows from G\"odel's original coordinates.
Note the opposite sign in the $g_{t\phi}$ component with respect to Refs.~\cite{Novello:1982nc,Iyer:1993qa,Rindler:1990}, where different notation for the overall scaling factor is also used.
Another form of the G\"odel metric with polynomial rather than hyperbolic function coefficients is obtained by the radial coordinate transformation $R/R_G=\sinh r$, where $R_G$ is a scaling constant with the dimensions of length \cite{Grave:2009zz}, such that $R<R_G$ (corresponding to $r<r_h$) in order to preserve the standard metric signature. 
}
\begin{eqnarray}
\label{met_godel}
d s^2&=&\frac{2}{\omega^2}\left[-d t^2 +d r^2 +\sinh^2 r(1-\sinh^2 r)\,d \phi^2 \right.
\nonumber\\
&&\qquad \left. +2\sqrt{2}\sinh^2r\, d t\,d \phi  +d z^2\right]
\nonumber\\
&=&\frac{2}{\omega^2}\left[-(d t -\sqrt{2}\sinh^2 r\,d\phi)^2 +d z^2 \right.
\nonumber\\
&&\qquad \left.  +d r^2 +\sinh^2 r(1+\sinh^2 r)\,d \phi^2\right]
\,,
\end{eqnarray}
where the latter form is the orthogonal decomposition adapted to the time coordinate lines.
The coordinates here are all dimensionless, while the scaling factor $\omega$ has the dimension of inverse length. 

The matter source for the G\"odel solution is a dust described by the stress-energy tensor  $T=\rho\, u\otimes u$, with constant energy density $\rho$ and unit 4-velocity $u=(\omega/\sqrt{2})\partial_t$ of the fluid particles aligned with the time coordinate lines, where  $\omega>0$ denotes the constant rotation parameter chosen to be positive to describe an intrinsic counterclockwise rotation of the spacetime of the particle world lines around the $z$-axis (increasing values of $\phi$). The cosmological constant has the value $\Lambda=-\omega^2=-4\pi\rho$.

The horizon radius
\beq
r_h=\ln (1+\sqrt{2})\approx 0.88137\,,
\eeq 
satisfying
\beq
\sinh r_h=1\,,\quad 
\cosh r_h=\sqrt{2}\,,
\eeq
is defined by the condition $ g_{\phi\phi}=0$,  where the $\phi$ coordinate circles are null and beyond which they are timelike. This radius delimits the ``physical region" of the spacetime for the given system of coordinates, if one wants to avoid closed timelike curves.

In the metric \eqref{met_godel} the time lines  are always timelike geodesics, so the coordinates are valid at all radii, but the spacelike $t=$constant hypersurfaces used to introduce observers (with 4-velocity $n$) along their normal direction turn timelike beyond $r_h$, so this represents an observer horizon for this family of fiducial observers. 
Their 4-velocity is
\begin{eqnarray}
    n&=& \frac{\omega}{\sqrt{2}} \frac{\sqrt{1-\sinh^2r}}{\cosh r} \left(\partial_t -\frac{\sqrt{2}}{1-\sinh^2 r} \partial_\phi\right) \,,\nonumber\\ 
n^\flat&=& -\frac{\sqrt{2}}{\omega}dt \,,\qquad r<r_h\,,
\end{eqnarray}
where the symbol $\flat$ denotes the fully covariant form a tensor.
The matter 4-velocity $u$ and its covariant form are 
\beq
\label{uflatgodel}
u=\frac{\omega}{\sqrt{2}}\partial_t\,,\qquad
u^\flat=-\frac{\sqrt{2}}{\omega}\left(d t -\sqrt{2}\sinh^2 r\, d \phi \right)\,.
\eeq 
This is  geodesic and shear-free but has a (constant) nonzero vorticity vector aligned with the cylindrical symmetry axes
\beq
\omega(u)=\frac{\omega^2}{\sqrt{2}}\partial_z\,,\qquad \omega(u)^\flat= \sqrt{2}\, d z\,.
\eeq
We will refer to these of observers as static, since they are at rest with respect to the spatial coordinates.
A spatial frame adapted to these observers is given by
\begin{eqnarray}
e(u)_1&=&\frac{\omega}{\sqrt{2}}\partial_r \,,\nonumber\\
e(u)_2&=&\frac{\omega}{\cosh r}\left[\sinh r\, \partial_t 
       +\frac{1}{\sqrt{2}\sinh r}\partial_\phi \right]\,,\nonumber\\ 
e(u)_3&=&\frac{\omega}{\sqrt{2}}\partial_z \,.
\end{eqnarray}
Hereafter, we will limit our considerations to the physical region $r<r_h$.

In order to discuss spin precession, it is useful to introduce the Cartesian-like frame in which the azimuthal rotation with $\phi$ is removed
\beq\label{exey}
  (e_x\ e_y) = (e(u)_1\ e(u)_2)\, \begin{pmatrix}\cos\phi&\sin\phi\\ -\sin\phi & \cos\phi\end{pmatrix}
\,.
\eeq

\subsection{Geodesics}

Because of the existence of the three Killing vectors fields $\partial_t$, $\partial_\phi$ and $\partial_z$, the geodesic equations are separable and the covariant 4-velocity of a general geodesic has the following  form, where $\lambda$ is an affine parameter
\beq
U^\flat=-E\,d t +L\,d \phi +b\,d z +U_r\, d r\,,
\eeq
with
\beq
U_r=\frac{2}{\omega^2}U^r=\frac{2}{\omega^2}\frac{dr}{d\lambda}\,,
\eeq
and normalization condition $U^\alpha U_\alpha=-\mu^2$, with $\mu^2=1,0,-1$ for timelike, null, spacelike geodesics, respectively.
The geodesic equations expressed in terms of the constants of the motion
$E=-U\cdot \partial_t$, $L=U\cdot \partial_\phi$ and $b=U\cdot \partial_z$  are
\begin{eqnarray}
\frac{dt}{d\lambda}&=&-\frac12\omega^2\left(E-\frac{\sqrt{2}X}{\cosh^2r}\right)
\,,\nonumber\\
\left(\frac{dr}{d\lambda}\right)^2&=&-\frac12\omega^2\mu^2-\frac14\omega^4(b^2+E^2)\nonumber\\
&&+\frac14\omega^4\frac{X^2}{\cosh^2r}-\frac14\omega^4\frac{L^2}{\sinh^2r}
\,,\nonumber\\
\frac{d\phi}{d\lambda}&=&-\frac12\frac{\omega^2}{\sinh^2r}\left(\sqrt{2}E-\frac{X}{\cosh^2r}\right)
\,,\nonumber\\
\frac{dz}{d\lambda}&=&\frac12\omega^2b
\,,
\end{eqnarray}
where $X=\sqrt{2}E+L$ and the separability of geodesics implies that one can find a coordinate system adapted to the geodesics \cite{Bini:2014uua}.

Introducing the nonnegative radial variable $x=\cosh^2r\ge1$ with horizon value $x_h=2$, the above equations become
\begin{eqnarray}
\label{geo_eqs}
\frac{dt}{d\lambda}&=&-\frac12\omega^2\left(E-\frac{\sqrt{2}X}{x}\right)
\,,\nonumber\\
\left(\frac{dx}{d\lambda}\right)^2&=&-\omega^2(2\mu^2+b^2\omega^2+\omega^2E^2)x^2-\omega^2(\omega^2L^2-2\mu^2\nonumber\\
&&-\omega^2X^2-\omega^2E^2-b^2\omega^2)x-\omega^4X^2\nonumber\\
&\equiv&Ax^2+Bx+C
\,,\nonumber\\
\frac{d\phi}{d\lambda}&=&-\frac12\frac{\omega^2}{x-1}\left(\sqrt{2}E-\frac{X}{x}\right)
\,,\nonumber\\
\frac{dz}{d\lambda}&=&\frac12\omega^2b
\,,
\end{eqnarray}
which maps the radial motion onto a quadratic potential well classical mechanical system with sinusoidal oscillations about its minimum.
The equation for $x$ can be rewritten as
\beq
\label{eqx}
\frac{dx}{d\lambda}=\epsilon_x \sqrt{|A|}\sqrt{(x-x_-)(x_+-x)}\,,
\eeq
with $\epsilon_x=\pm1$ keeping track of increasing/decreasing radial coordinate, and
\begin{eqnarray}
x_\pm&=&1+\frac{\omega^4}{A}(L^2-X^2) \\
&&\mp\frac{1}{2A}\left[(\omega^4(X-L)^2+A)(\omega^4(X+L)^2+A)\right]^{1/2}\,,\nonumber
\end{eqnarray}
are the turning points of the radial motion, which is confined to the region $x_-\leq x\leq x_+$.
Both the $t$ and $\phi$ motions have turning points, located  at 
\beq
\label{tptandphi}
x_t=\frac{\sqrt{2}X}{E}\,,\qquad
x_\phi=\frac{X}{\sqrt{2}E}
=\frac12x_t
\,,
\eeq
respectively, as it follows immediately from Eqs.~\eqref{geo_eqs}.

A careful study of general geodesic motion using the effective potential approach can be found in Ref.~\cite{Novello:1982nc}, which derives their analytic solution. 
In fact, the radial equation \eqref{eqx} can be immediately integrated to yield
\beq
\epsilon_x \sqrt{|A|}\lambda=\arcsin\left(\frac{2x-x_+-x_-}{x_+-x_-}\right)\,,
\eeq
choosing $\lambda=0$ to correspond to $x=\bar x=\frac12(x_++x_-)$.
This relation can then be easily inverted
\beq
x(\lambda)=-\epsilon_x\delta\sin(\sqrt{|A|}\lambda)+\bar x\,, \qquad
\delta=\frac12(x_+-x_-)\,.
\eeq
One then finds
\begin{widetext}
\begin{eqnarray}
t(\lambda)
&=&\epsilon_x\frac{\sqrt{2}\omega^2X}{\sqrt{|A|x_+x_-}}\left[{\rm arctan}\left(\frac{\epsilon_x\bar xT(\lambda)+\delta}{\sqrt{x_+x_-}}\right)
-{\rm arctan}\left(\frac{\delta}{\sqrt{x_+x_-}}\right)\right]-\frac12\omega^2E\lambda
\,,\nonumber\\
\phi(\lambda)
&=&-\epsilon_x\frac{\omega^2(\sqrt{2}E-X)}{\sqrt{|A|(x_+-1)(x_--1)}}\left[
{\rm arctan}\left(\frac{\epsilon_x(\bar x-1)T(\lambda)+\delta}{\sqrt{(x_+-1)(x_--1)}}\right)
-{\rm arctan}\left(\frac{\delta}{\sqrt{(x_+-1)(x_--1)}}\right)
\right]\nonumber\\
&&
-\epsilon_x\frac{\omega^2X}{\sqrt{|A|x_+x_-}}\left[
{\rm arctan}\left(\frac{\epsilon_x\bar xT(\lambda)+\delta}{\sqrt{x_+x_-}}\right)-{\rm arctan}\left(\frac{\delta}{\sqrt{x_+x_-}}\right)
\right]
\,,\nonumber\\
z(\lambda)&=&\frac12\omega^2b\lambda
\,,
\end{eqnarray}
\end{widetext}
where 
\beq
T(\lambda)=\tan\left(\frac{\sqrt{|A|}\lambda}{2}\right)\,,
\eeq
and initial conditions have been chosen so that $t(0)=\phi(0)=z(0)=0$.
We are interested in the orbits $b=0$ which are confined to ``planes" of constant $z$ orthogonal to the cylindrical symmetry axes.

All orbits have been classified in Ref.~\cite{Novello:1982nc} according to the ratio between angular momentum and energy. 
We investigate below the features of elliptic-like geodesics on constant $z$ hypersurfaces by using a $(p,e)$ orbit parametrization  familiar from Newtonian mechanics,  where $p$ and $e$ are the semi-latus rectum and eccentricity, respectively.

\section{Elliptic-like ``planar" geodesics}

Let us consider timelike geodesics ($\mu=1$)  on a constant $z$ hyperplane, i.e., with $b=0$. The affine parameter is then proper time $\lambda=\tau$.
The associated 4-velocity is
\begin{eqnarray}
   U &=& \frac{\omega^2}{2\cosh^2 r} \left[ \sqrt{2} L + E (2-\cosh^2 r) \right] \,\partial_t 
       +U^r \partial_r 
\nonumber\\
&& 
- \frac{\omega^2}{2 \sinh^2 r \cosh^2 r} \left( E \sqrt{2} \sinh^2 r - L\right) \,\partial_\phi
\,,
\end{eqnarray}
where 
\begin{eqnarray}
\label{Urequat}
(U^r)^2&=& \frac{\omega^2}{4\sinh^2 r \cosh^2 r}\left[-2(\tilde E^2+1)\cosh^4 r \right.\\
&+& \left. 2(3\tilde E^2+2\tilde E \tilde L+1)\cosh^2 r-(2\tilde E+\tilde L)^2  \right]
\,,\nonumber
\end{eqnarray}
having introduced the following rescaling of $E$ and $L$
\beq
\tilde E=\frac{\omega}{\sqrt{2}}  E\ge1\,, \qquad
\tilde L=\omega L\,.
\eeq
The square bracketed expression in $U^r$ is quadratic in $x=\cosh^2 r$ with real factors which determine the turning points of the radial motion 
\begin{eqnarray}
\label{Urfactorized}
|U^r|&=& \frac{\omega \sqrt{2}\sqrt{\tilde E^2+1}}{2\sinh r \cosh r} \\
&\times&\sqrt{(\cosh^2r-\cosh^2r_{\rm (peri)})(\cosh^2r_{\rm (apo)}-\cosh^2 r)},\nonumber
\end{eqnarray}
where 
\begin{eqnarray}
\label{rootsUr}
\cosh^2r_{\rm (peri)}&=&  
1+\frac{ \tilde E^2-1+2\tilde L\tilde E - \sqrt{\Delta} }{2 (\tilde E^2+1)} 
\,,\nonumber\\
\cosh^2r_{\rm (apo)}&=& 
1+\frac{ \tilde E^2-1+2\tilde L\tilde E + \sqrt{\Delta} }{2 (\tilde E^2+1)} 
\,,
\end{eqnarray}
 with
\beq
 \Delta = (\tilde E^2-1)(\tilde E^2+4\tilde L\tilde E+2\tilde L^2-1)\geq0\,.
\eeq

The conditions for the existence of the roots $r_{\rm (peri/apo)}$ are given by
\beq
\label{condsrperiapo}
\Delta\geq0\,,\qquad
\tilde E^2-1+2\tilde L\tilde E \pm \sqrt{\Delta}\geq0\,. 
\eeq
$\Delta$ vanishes identically for $\tilde E=1$ and every value of $\tilde L$, and along the curves
\beq
\label{zeros_Delta0}
\tilde L =-\tilde E\pm \frac{1}{\sqrt{2}}\sqrt{\tilde E^2+1}\equiv \tilde L^{(\Delta)}_{\pm} (\tilde E)\,.
\eeq
Further requiring that $r_{\rm (apo)}<r_h$, i.e., $\cosh^2r_{\rm (apo)}<2$, the second condition in Eq.~\eqref{condsrperiapo} implies
\beq
\tilde L<2(\tilde E-\sqrt{\tilde E^2-1})\equiv \tilde L_h(\tilde E)\,.
\eeq 
The case $\tilde L=0$ should be treated separately.
In fact, in this case $\Delta = (\tilde E^2-1)^2$, so that 
\begin{eqnarray}
\cosh^2r_{\rm (peri)}&=&1+\frac{ \tilde E^2-1 - |\tilde E^2-1| }{2 (\tilde E^2+1)} 
\,,\nonumber\\
\cosh^2r_{\rm (apo)}&=&1+\frac{ \tilde E^2-1 + |\tilde E^2-1| }{2 (\tilde E^2+1)} 
\,.
\end{eqnarray}
Hence, if $\tilde E>1$ then $\cosh^2r_{\rm (peri)}=1$ so that the orbit passes through the origin  and $\cosh^2r_{\rm (apo)}=2\tilde E^2/(\tilde E^2+1)$.
The case  $\tilde E<1$ instead must be discarded, since it would imply $\cosh^2r_{\rm (peri)}=2\tilde E^2/(\tilde E^2+1)<1$ and $\cosh^2r_{\rm (apo)}=1$. 

\begin{figure*}
\begin{center}
$\begin{array}{cc}
\includegraphics[scale=0.3]{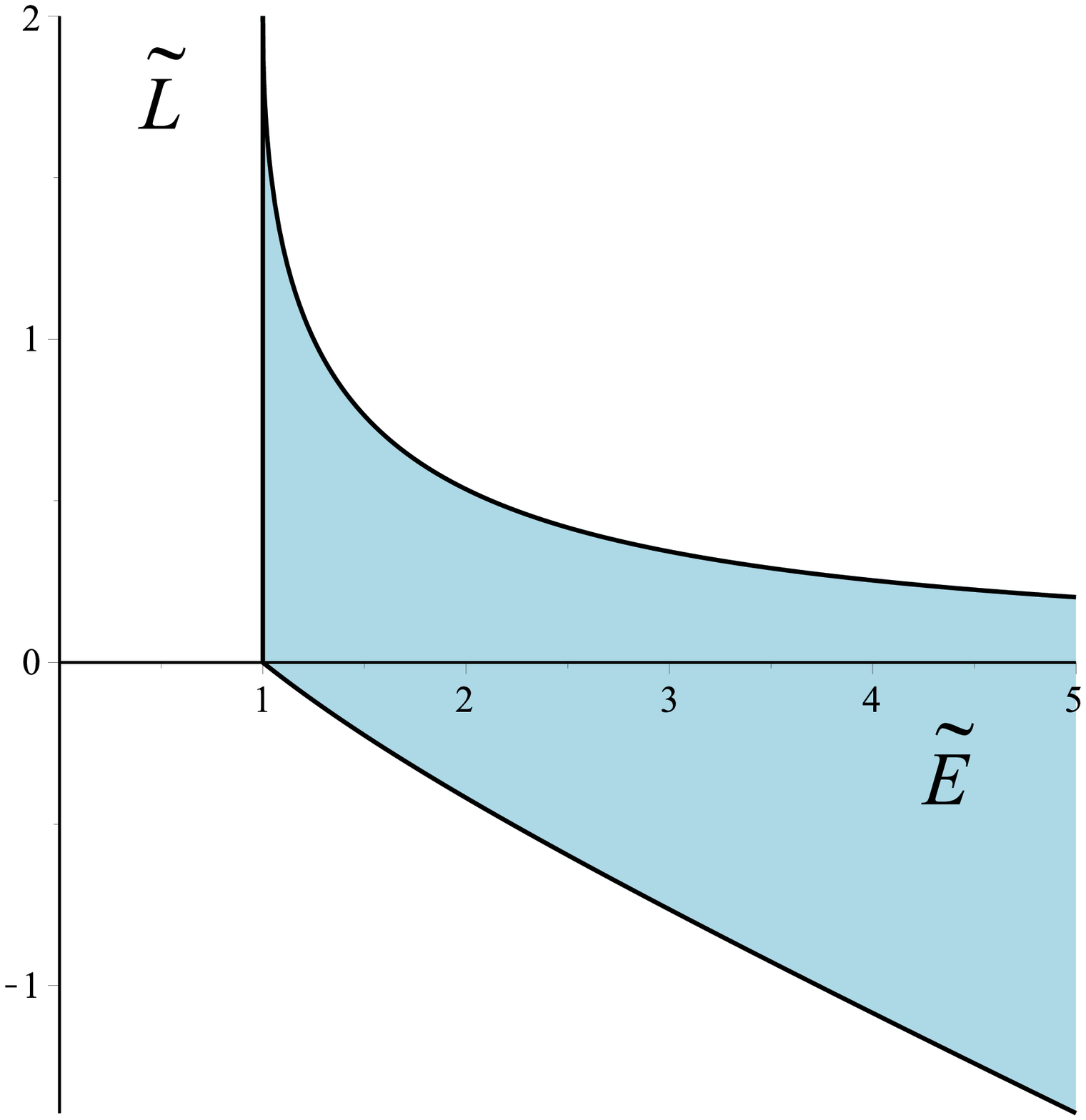}&\qquad
\includegraphics[scale=0.3]{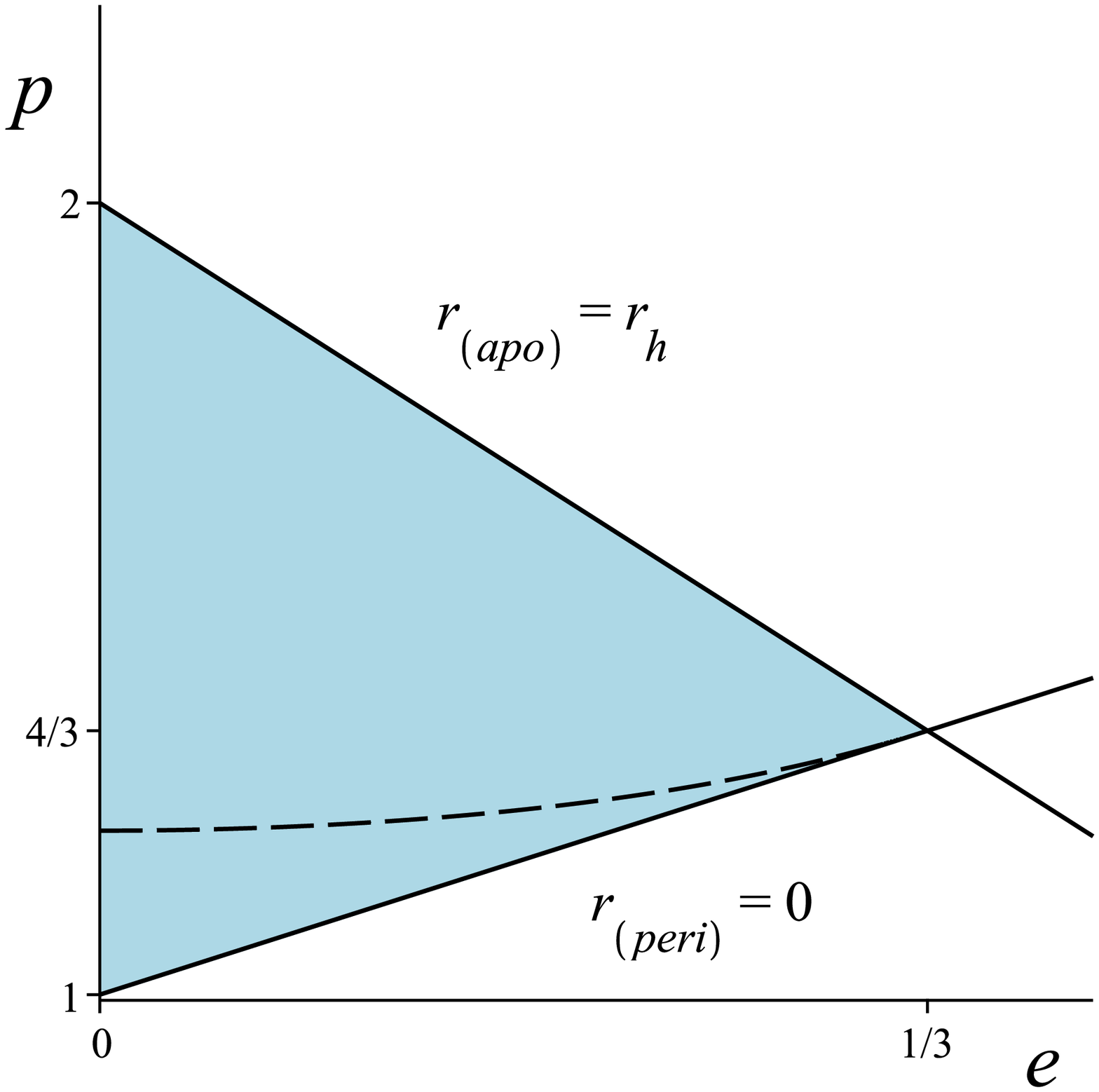}\\[.2cm]
\mbox{(a)} &\qquad \mbox{(b)}\cr
\end{array}
$\\
\end{center}
\caption{\label{fig:1} The allowed parameter spaces $(\tilde L,\tilde E)$ and $(p,e)$ for timelike geodesics in the physical region $r<r_h$ are shown in panels (a) and (b), respectively (shaded regions). In (a) the sign of $\tilde L$ determines the 3 types of allowed orbits within the angular momentum bounding curves: type I below the axis, type II along the axis and type III above the axis.
Panel (b) shows the corresponding allowed region of the $(p,e)$ parameter space bounded below by the line $p=1+e$ of the type II orbits where $r_{\rm (peri)}=0$ and $\tilde L=0$ and above by the line $p=2(1-e)$ where $r_{\rm (apo)}=r_h$ and $\tilde L=\tilde L_h$ (see Eq.~\eqref{rpmdef}). The type I and III orbits are folded over onto each other into the triangular region across the type II boundary.
Type I orbits are confined to the region below the dashed curve, which is implicitly defined by the vanishing of the denominator in Eq.~\eqref{tilde_E_L} and corresponds to null geodesics.
}
\end{figure*}

Summarizing, the allowed range of $\tilde L$ as a function of $\tilde E$ is given by
\beq
\tilde L_{\rm min}(\tilde E)\leq \tilde L(\tilde E)<\tilde L_h(\tilde E)\,,
\eeq
where $\tilde L_{\rm min}(\tilde E)\equiv \tilde L^{(\Delta)}_{+}(\tilde E)\leq0$ for every value of $\tilde E\geq1$.
The allowed parameter space $(\tilde L,\tilde E)$ is shown in Fig.~\ref{fig:1} (a).

The orbits can then be classified according to whether the angular momentum is positive, negative or zero for fixed values of the energy parameter \cite{Novello:1982nc}. Turning points for either the $t$ and $\phi$ motions may occur at  (see Eq.~\eqref{tptandphi})
\begin{eqnarray}
\label{tptandphi2}
\cosh^2r_t &=& 2+\frac{\tilde L}{\tilde E}\,,\nonumber\\
\cosh^2r_\phi &=& 1+\frac{\tilde L}{2\tilde E}
\,,
\end{eqnarray}
respectively.
We will refer below to orbits with $\tilde L_{\rm min}\leq\tilde L<0$ as type I, orbits with $\tilde L=0$ as type II, and orbits with $0<\tilde L< L_h$ as type III. Their main properties are summarized as follows.

\begin{enumerate}

\item $\tilde L_{\rm min}\leq\tilde L<0$ (type I):

The orbits have $0<r_{\rm (peri)}\leq r\leq r_{\rm (apo)}$ and go around the origin opposing the local rotation of the dust source with no turning point in $\phi$ (which is monotonically decreasing\footnote{
This sign correlation is due to our choice of the sign in the $g_{t\phi}$ component of the spacetime metric.
In fact, reversing this sign makes $\phi$ monotonically increase (see, e.g., Ref.~\cite{Novello:1982nc}).%
}%
) and $t$ ($r_t>r_{\rm (apo)}$).
The orbit becomes circular for $r_{\rm (peri)}=r_{\rm (apo)}$.

\item $\tilde L=0$ (type II):

The same as type I, with $r_{\rm (peri)}=0$, and $\cosh^2r_{\rm (apo)}=2\tilde E^2/(\tilde E^2+1)$, so that the orbits pass through the origin.

\item $0<\tilde L< L_h$ (type III):

In this case $r_{\rm (peri)}\leq r_\phi\leq r_{\rm (apo)}$, whereas $r_t>r_{\rm (apo)}$, so that the $\phi$-motion has a turning point in the allowed range of $r$, implying that the orbit cannot go around the origin.
No circular orbits exist.

\end{enumerate}

\begin{figure*}
\includegraphics[scale=1]{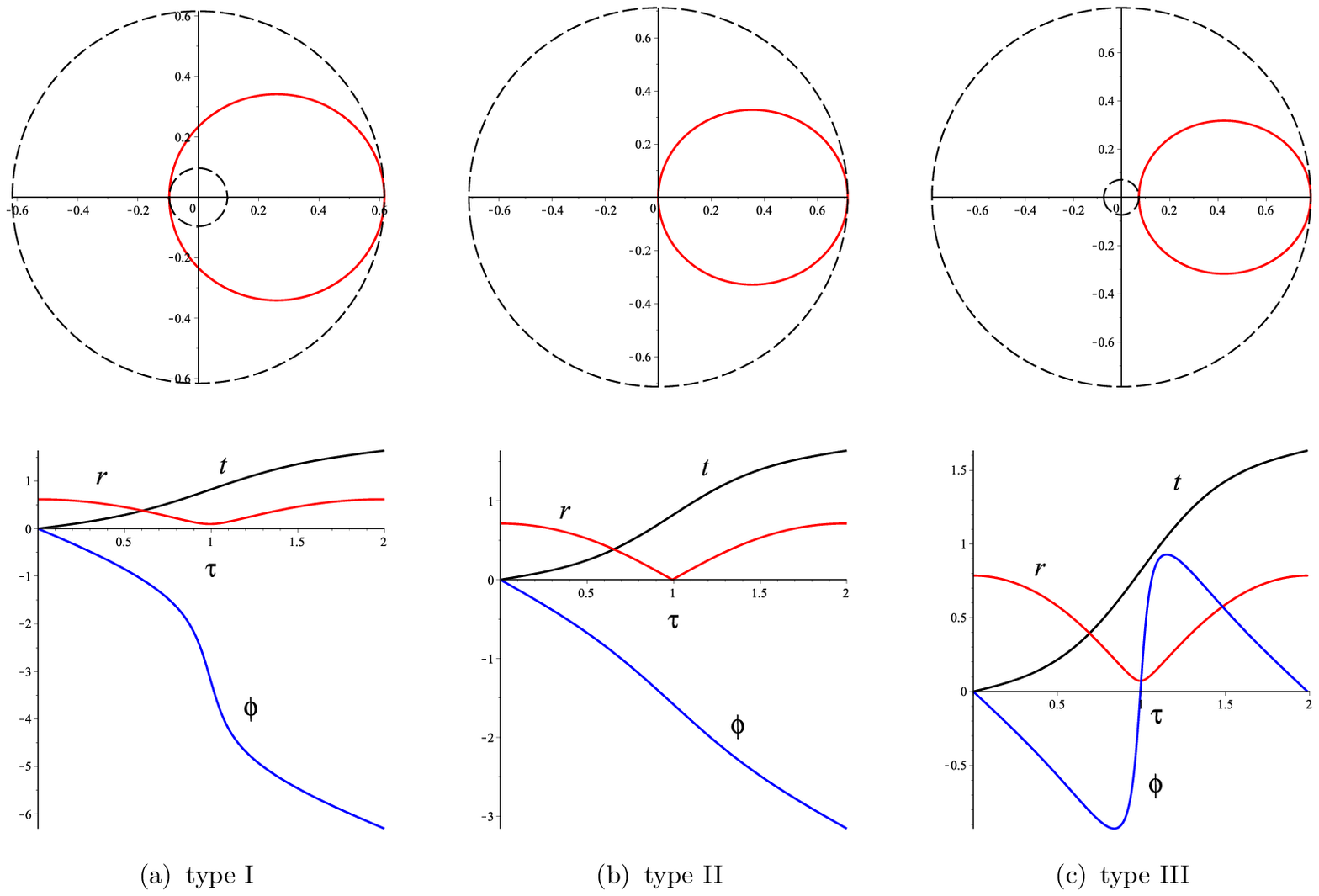}
\caption{
Various kinds of timelike geodesic planar motion orthogonal to the symmetry axes are shown, together with circles indicating the periastron and apoastron radii. The orbits are integrated numerically for the parameter choice $\omega=1$, $\tilde E=2$ and the selected values $\tilde L=[-0.2,0,0.2]$ for orbits of types I, II and III, respectively. The corresponding values of $(p,e)$ are $p\approx1.1836$ and $e\approx0.1727$ (type I), $p=1+e$ and $e\approx0.2308$ (type II), and $p\approx1.2783$ and $e\approx0.2715$ (type III).
The initial conditions are chosen so that $r(0)=r_{\rm (apo)}\approx[0.6165,0.7127,0.7854]$, $t(0)=0$, $\phi(0)=0$. Each of the loops is traversed clockwise (see footnote 2).
}
\label{fig:2}
\end{figure*}

Some examples of numerically integrated orbits are shown in Fig.~\ref{fig:2}. In all three cases the motion around the orbit loop is in the clockwise direction, opposing the counterclockwise local rotation of the dust source world lines. This is the same behavior exhibited by timelike geodesics in flat spacetime in a rotating cylindrical coordinate system: straight line radial outward geodesic paths appear to rotate backwards with respect to the forward rotation of the cylindrical coordinates.

Note that while all timelike geodesics have circular orbits with respect to cylindrical coordinates based on the center of those ``circles," the transformation between different such cylindrical coordinate systems is nontrivial, each involving distinct time coordinate slicings of the spacetime.
Thus comparing results for distinct geodesics among the three different classes of orbits in a single cylindrical coordinate system is also not trivial. Indeed these curves are helices in the spacetime, and the precession with respect to one cylindrical coordinate coordinate grid as a function of proper time along that helix is quite different along another one, since the radial and polar angle grid directions in the quotient space by the time lines must be distorted in the change of coordinate systems in a way that does not allow comparison of equal increments in the angular coordinates. However, the total precession angle around one loop which begins and returns to the same dust particle world line is independent of the coordinate system (see Appendix A for details).


\subsection{Newtonian-like parametrization of the orbits}

Although the new radial variable $x$ is simply related to the proper time parametrization because of its quadratic potential well energy formulation, the other variables and the dependence on the constants of the motion  $(\tilde E,\tilde L)$ is very complicated involving many square root expressions. Drawing on the analogy with the classical Kepler problem where $1/r$ has quadratic potential well dynamics when parametrized by the polar angle, leading to the usual conic section parametrization of $r$ in terms of that angle, we can introduce a similar conic section angle parametrization for $x$ here
\beq
\label{rdichirel}
x=\cosh^2 r=\frac{p}{1+e\cos\chi}\,.
\eeq
Here $p$ and $e$  represent the semi-latus rectum and eccentricity, respectively, in analogy with a Newtonian elliptical orbit and polar angle $\chi$ in the plane of the orbit where $x$ and $\chi$ are interpreted as ordinary polar coordinates related to the actual polar coordinates by the transformation 
$(r,\phi)=({\rm arccosh}\, \sqrt{x},\phi(\chi))$, but interpreting this as a change of radial variable from $r$ to $\chi$ along each geodesic depending on the conserved parameters, and using that new variable to parametrize the orbit in spacetime.
The allowed conserved quantity parameter ranges are $p\geq1$ and as usual for elliptical orbits, $0\leq e<1$,  so that $r=r_{\rm (peri)}$ for $\chi=0$ and $r=r_{\rm (apo)}$ for $\chi=\pi$, i.e., 
\beq
\label{rpmdef}
\cosh^2r_{\rm (peri)} = \frac{p}{1+ e}\,,\qquad
\cosh^2r_{\rm (apo)} = \frac{p}{1- e}\,.
\eeq

\subsubsection{Parameter space}

The conditions for the existence of the roots \eqref{rpmdef} in the physical region $r<r_h$ are simply given by
\beq
\label{epparspace}
\frac{p}{1+ e}\geq1\,,\qquad 
\frac{p}{1- e}<2\,.
\eeq
The allowed  $(p,e)$ parameter space is shown in Fig.~\ref{fig:1} (b). 

Direct comparison between Eqs.~\eqref{rootsUr} and \eqref{rpmdef}, or equivalently imposing the conditions $U^r(r_{\rm (peri)})=0=U^r(r_{\rm (apo)})$ by using Eq.~\eqref{Urfactorized}, allows one to express $\tilde E$ and $\tilde L$ as functions of $p$ and $e$
\begin{eqnarray}
\label{tilde_E_L}
\tilde E_\pm &=& \frac{p\pm Y}{\sqrt{(p\mp Y)^2-4p(p-1)}}
\,,\nonumber\\
\tilde L_\pm &=& \frac{\mp2Y}{\sqrt{(p\mp Y)^2-4p(p-1)}}\,,
\end{eqnarray}
where $Y=\sqrt{(p-1)^2-e^2}$ and the $\pm$ signs here refer to the two branches of $\tilde E$ and $\tilde L$.
The argument of the square root in the denominator is always positive in the allowed region for $(p,e)$ if the $-$ sign is selected (i.e., for $(\tilde E_+,\tilde L_+)$).
In contrast, requiring its positiveness with the $+$ sign (i.e., for $(\tilde E_-,\tilde L_-)$) gives a restriction on the parameters (see Fig.~\ref{fig:1}) which prevents the timelike geodesics from turning null when it vanishes.

Equation \eqref{tilde_E_L} implies the following relation between $\tilde E$ and $\tilde L$
\beq
\label{relELpm}
\frac{\tilde L_\pm}{\tilde E_\pm}=\frac{-2}{1\pm\displaystyle\frac{p}{\sqrt{(p-1)^2-e^2}}}\,.
\eeq
The $+$ solution corresponds to negative values of $\tilde L$, and hence to orbits of type I, including the circular case, whereas the $-$ solution corresponds to positive values of $\tilde L$, and hence to orbits of type III.
Finally, type II orbits correspond to $\tilde L=0$, i.e., $p=1+e$ and $e<1/3$, with $\tilde E=\sqrt{{(1+e)}/{(1-3e)}}$.
In any case the eccentricity is limited to $e<1/3$.

The following useful relations can be simply obtained from Eqs.~\eqref{tilde_E_L} and \eqref{relELpm}
\begin{eqnarray}
\label{usefulrel}
\tilde L_\pm&=&\mp\frac{2\zeta_\pm p}{\omega}\frac{\sqrt{1-e^2}}{\sqrt{(p-1)^2-e^2}}
\,,\nonumber\\
\tilde L_\pm+2\tilde E_\pm&=&\mp\frac{p\tilde L_\pm}{\sqrt{(p-1)^2-e^2}}=\frac{2\zeta_\pm p}{\omega\sqrt{1-e^2}}\,,
\end{eqnarray}
where $\zeta_\pm$ is defined in Eq.~\eqref{zetadef} with $\tilde E=\tilde E_\pm$.

\subsection{Equations of motion}

The equation for the radial variable $r$ is thus replaced by the following equation for the angular variable $\chi$ whose velocity undergoes a simple harmonic velocity oscillation in proper time $\tau$
\beq
\frac{d\chi}{d \tau}=\frac{\omega\sqrt{2}\sqrt{\tilde E^2+1}}{\sqrt{1-e^2}}(1+e\cos \chi)\,, 
\eeq
whose integration gives
\begin{eqnarray}
&&\frac{2}{\sqrt{1-e^2}}{\rm arctan}\left(\sqrt{\frac{1-e}{1+e}} \tan \frac{\chi}{2}
\right)\nonumber\\
&&\qquad= \frac{\omega  \sqrt{2}\sqrt{\tilde E^2+1}}{\sqrt{1-e^2}}  \tau + {\rm const.}
\end{eqnarray}
Choosing the initial condition $\chi(0)=0$ and defining
\beq
\label{zetadef}
\zeta =  \frac{ \omega\sqrt{\tilde E^2+1}}{  \sqrt{2}}\,,
\eeq
we find
\begin{eqnarray}
\label{chi_sol_fin}
 \tan \frac{\chi}{2}
&=&\sqrt{\frac{1+e}{1-e}}\tan \left(\zeta \tau  \right)\,, \nonumber\\
\tau &=& \frac{1}{\zeta}{\rm arctan}\left( \sqrt{\frac{1-e}{1+e}} \tan \frac{\chi}{2}\right)
\,.
\end{eqnarray}
The evolution equations for $t$ and $\phi$ can then be rewritten as
\begin{eqnarray}
\frac{dt}{d\chi} &=&\frac{(1-e^2)^{1/2}}{2p(\tilde E^2+1)^{1/2}}\left(
\tilde L+2\tilde E-\frac{p \tilde E}{1-e\cos\chi}
\right)
\,,\nonumber\\
\frac{d\phi}{d\chi} &=&-\frac{\sqrt{2} (1-e^2)^{1/2}}{4 p (\tilde E^2+1)^{1/2}}\left( 
\tilde L+2\tilde E +\frac{p \tilde L}{1-p+e\cos\chi} 
\right)
\,,\nonumber\\
\end{eqnarray}
with solution
\begin{eqnarray}
\label{soltphidichi}
t(\chi) &=&\frac{\omega}{\sqrt{2}\zeta} \left[\frac{\sqrt{1-e^2}}{2p}(\tilde L+2\tilde E)\,\chi-\tilde E\,{\rm arctan} (\psi_1) \right]
\,,\nonumber\\
\phi(\chi) &=&-\frac{\omega \sqrt{1-e^2}}{2\zeta} \left[\frac{1}{2p}(\tilde L+2\tilde E)\,\chi \right. \nonumber\\
&&\left. -  \frac{\tilde L}{\sqrt{(p-1)^2-e^2}}  {\rm arctan} (\psi_2)\right]\,,
\end{eqnarray}
where
\beq
\psi_1= \sqrt{\frac{1-e}{1+e}}\tan \frac{\chi}{2}\,,\qquad 
\psi_2= \sqrt{\frac{p-1+e}{p-1-e}} \tan \frac{\chi}{2}\,.
\eeq

\subsubsection{Periods and frequencies}

The coordinate time period $T_r$ of the radial motion and the corresponding full variation $\Phi$ of the azimuthal angle are given by
\begin{eqnarray}
\label{periods}
T_r&=&\oint dt = 2 \int_0^\pi \frac{dt}{d\chi} d\chi\nonumber\\
&=&
\frac{\omega\pi\sqrt{1-e^2}}{\sqrt{2}\zeta p} \left[\tilde L+2\tilde E-\frac{\tilde E p}{\sqrt{1-e^2}}\right]
\,,\nonumber\\
\Phi &=&\oint d\phi  = 2 \int_0^\pi \frac{d\phi}{d\chi} d\chi\nonumber\\
&=&
-\frac{\omega\pi\sqrt{1-e^2}}{2\zeta p} \left[\tilde L+2\tilde E-\frac{\tilde Lp}{\sqrt{(p-1)^2-e^2}}\right]
\,.
\end{eqnarray}
The associated radial and azimuthal frequencies are given by $\Omega_r={2\pi}/{T_r}$ and $\Omega_\phi={\Phi}/{T_r}$, respectively.
The periastron advance is defined by
\beq\label{K}
K=\frac{|\Phi|}{2\pi}-1=\frac{|\Omega_\phi|}{\Omega_r}-1\,.
\eeq 
A method to visualize precession effects using the correspondence between the actual orbit and an instantaneous, fictitious one is discussed in Appendix B.

Finally, the proper time period ${\mathcal T}_r$ of the radial motion is given by
\begin{eqnarray}
\label{Taurdef}
{\mathcal T}_r&=&\oint d\tau  = 2 \int_0^\pi \frac{d\tau}{d\chi} d\chi
=
\frac{\pi}{\zeta}\,.
\end{eqnarray}
A related quantity is the redshift invariant 
\beq
z_1=\frac{{\mathcal T}_r}{T_r}\,,
\eeq
which has been studied extensively in the last few years in black hole perturbation theory within the self-force framework due to its gauge-invariant character (see, e.g., Refs.~\cite{Detweiler:2008ft,LeTiec:2011ab,Bini:2013zaa,Kavanagh:2015lva}).

We specialize below the previous expressions to orbits of each type.

\subsection{Newtonian-like elliptic orbits (type I)}

We are mostly interested here in Newtonian-like elliptic orbits going around the origin, i.e., orbits of type I with $\tilde E=\tilde E_+$ and $\tilde L=\tilde L_+$ in Eqs.~\eqref{tilde_E_L} and \eqref{relELpm}. 

The solutions \eqref{soltphidichi} for $t$ and $\phi$ become 
\begin{eqnarray}
\label{phidichitypeI}
t(\chi)&=&\frac{1}{\sqrt{2}}\left[\chi-\frac{\sqrt{(p-1)^2-e^2}+p}{\sqrt{1-e^2}} {\rm arctan} (\psi_1)\right]
\,,\nonumber\\
\phi(\chi) &=&-\frac{\chi}{2}-{\rm arctan} (\psi_2)\,,
\end{eqnarray}
where the simplifying relations \eqref{relELpm} and \eqref{usefulrel} have been used.
The behavior of $\phi$ as a function of $\chi$ is shown in Fig.~\ref{fig:phi_vs_chi} (a) for a selected value of $e$.

\subsubsection{Circular limit}

Along circular orbits ($e=0$) at a fixed coordinate radius the energy and angular momentum \eqref{tilde_E_L} and their ratio \eqref{relELpm} reduce to
\beq
\tilde E_{\rm(circ)} =\frac{2p-1}{{\mathcal D}(p)}\,,\qquad
\tilde L_{\rm(circ)}=-\frac{2(p-1)}{{\mathcal D}(p)}\,,
\eeq
and 
\beq
\frac{\tilde L_{\rm(circ)}}{\tilde E_{\rm(circ)}}=-\frac{2(p-1)}{2p-1}\,,
\eeq
respectively, where
\beq
\label{rstardef}
{\mathcal D}(p)=(-4p^2+4p+1)^{1/2}\,, 
\eeq
with
\beq
1\leq p<(1+\sqrt{2})/2\equiv \cosh^2r_* \ (\approx 1.20711)\,.
\eeq
The associated 4-velocity is given by
\begin{eqnarray}
U_{\rm (circ)}&=&\Gamma_{\rm (circ)} (\partial_t +\Omega_{\rm (circ)} \partial_\phi)\,,\nonumber\\
\Omega_{\rm (circ)} &=& -\frac{2^{3/2}}{3-2p}\,,\nonumber\\
\Gamma_{\rm (circ)} &=& \frac{\sqrt{2}\omega (3-2p)}{2 {\mathcal D}(p)}
\equiv\frac{\omega}{\sqrt{2}} \tilde\Gamma_{\rm (circ)}
\,.
\end{eqnarray}
Therefore, circular geodesics exist in the region $0<r<r_*$, with $r_*<r_h$. The boundary value $r=r_*$ is the light-ring, corresponding to $r=3M$ in analogous situation in the familiar case of a Schwarzschild black hole.

Expanding the exact relations \eqref{tilde_E_L} and \eqref{relELpm} in a series in $e$ gives
\begin{eqnarray}
\tilde E &=& \frac{2p-1}{{\mathcal D}(p)}+\frac{p(2p-3)}{(p-1){\mathcal D}^3(p)}e^2+O(e^4)\,, \nonumber\\
\tilde L &=& -\frac{2(p-1)}{{\mathcal D}(p)}-\frac{p(4p-5)}{(p-1){\mathcal D}^3(p)}e^2+O(e^4) \,,
\end{eqnarray}
and
\beq
\frac{\tilde L}{\tilde E}=-\frac{2(p-1)}{2p-1}+\frac{p}{(p-1)(2p-1)^2}e^2+O(e^4)\,,
\eeq
respectively.

\subsubsection{Periods and frequencies}

The coordinate time period $T_r$ of the radial motion and the period $\Phi$ of the azimuthal motion \eqref{periods} are given by
\begin{eqnarray}
\label{periods1}
T_r&=&\sqrt{2}\pi\left[1-\frac{p+\sqrt{(p-1)^2-e^2}}{2\sqrt{1-e^2}}\right]
\nonumber\\
&=&\frac{\pi}{\sqrt{2}}(3-2p)\left[1+\frac{p}{2(p-1)}\, e^2 +O(e^4)\right] 
\,,\nonumber\\
\Phi&=&-2\pi\,,
\end{eqnarray}
implying that there is no periastron advance ($K=0$) for these orbits, in contrast with the corresponding black hole case.
Furthermore, the associated radial and azimuthal frequencies are degenerate, namely $\Omega_r=-\Omega_\phi$, with
\beq
\Omega_r=\frac{2\pi}{T_r}=  \frac{2^{3/2}}{3-2p}+\frac{\sqrt{2} p(3-2p)}{(p-1)} e^2+O(e^4)\,.
\eeq

\begin{figure*}
\begin{center}
$\begin{array}{ccc}
\includegraphics[scale=0.25]{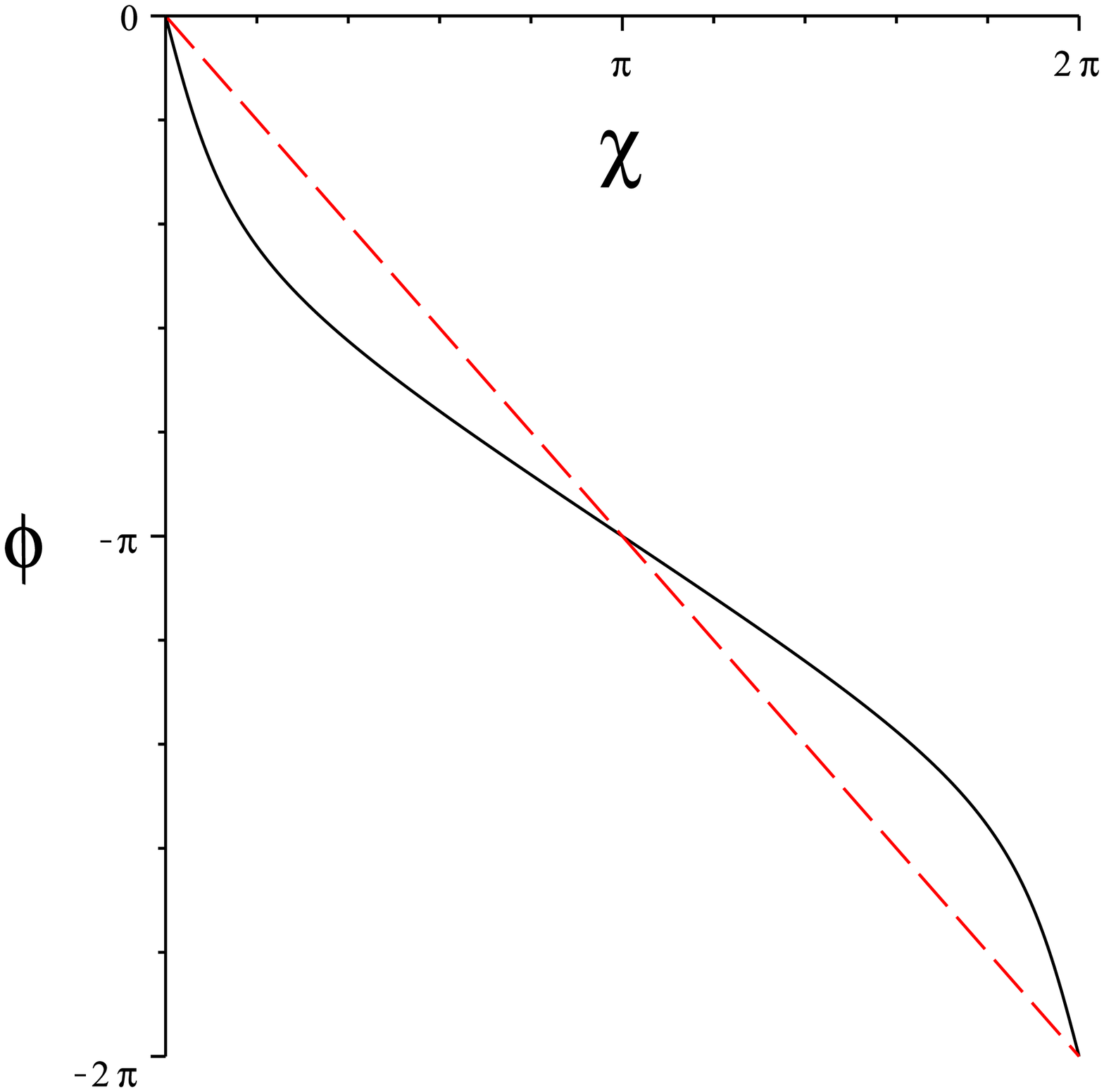}&\qquad
\includegraphics[scale=0.25]{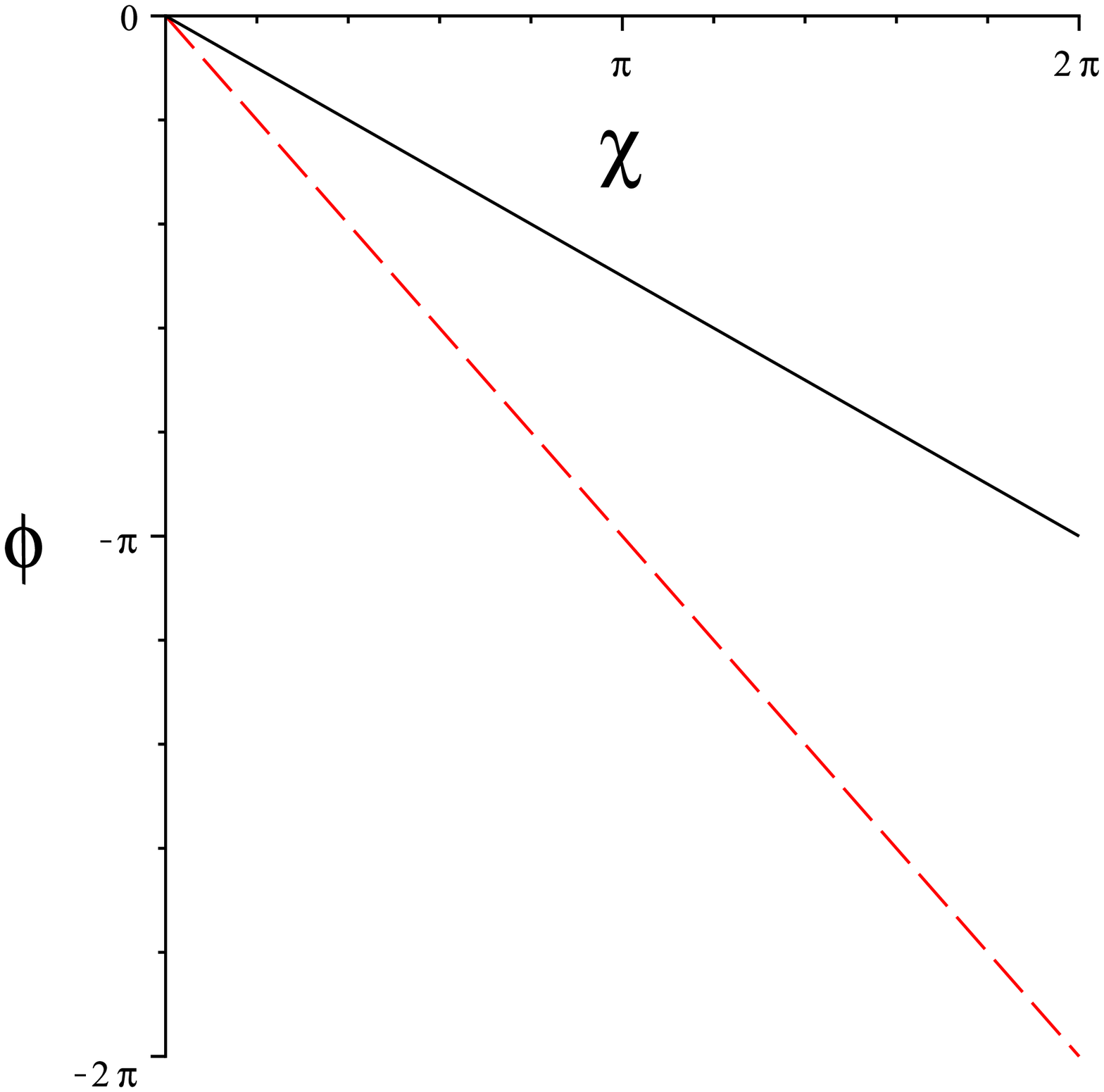}&\qquad
\includegraphics[scale=0.25]{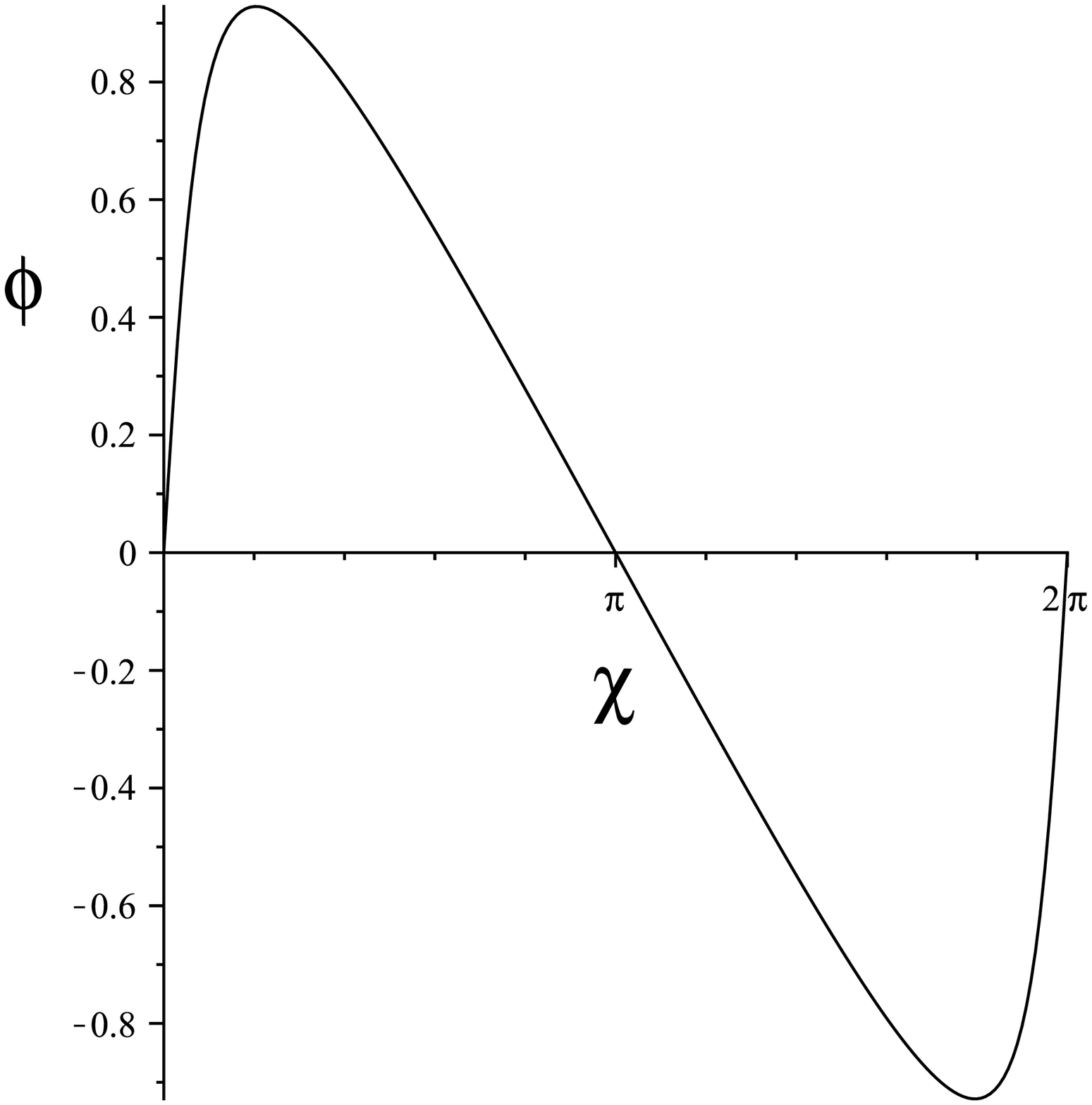}\\\\[.4cm]
\mbox{(a)\, type I} &\qquad \mbox{(b)\, type II} &\qquad \mbox{(c)\, type III}
\end{array}$\\
\end{center}
\caption{The behavior of $\phi$ as a function of $\chi$ is shown for the same choice of parameters as in Fig.~\ref{fig:2} (see Eqs. \eqref{phidichitypeI}, \eqref{phidichitypeII} and \eqref{phidichitypeIII} for orbits of type I, II and III, respectively).
}
\label{fig:phi_vs_chi}
\end{figure*}

Fig.~\ref{fig:phi_vs_chi} (a) shows how over the course of one period of the motion, the angle $\phi$ oscillates from the angle $\chi$ which describes fixed positions on the instantaneous $x$-ellipse orbit described in Appendix B whose precession traces out the actual orbit. When plotted together, the symmetry axes of this instantaneous orbit and of the actual orbit are aligned at apoastron and periastron, but departing from the periastron, the axes of the $x$-ellipse precess clockwise in the direction of the motion for a quarter cycle in $\phi$ to a maximum of about 45$^\circ$ for these values of the parameters and then precesses in the counterclockwise direction for half a cycle passing through alignment of the apoastron at midcycle, and then reverses direction again for the final quarter cycle returning to alignment at the periastron. Thus no net precession takes place during one orbit, in sharp contrast to the accumulated precession in a consistent direction which occurs for corresponding instantaneous ellipses in the case of bound equatorial plane orbits around a black hole. The precessing ellipses here ``wobble" back and forth in a single cycle returning to their original position after one period of the motion (see Fig.~\ref{fig:wobble} (a)).  Appendix B explains this instantaneous $x$-ellipse construction.

Finally, the proper time period ${\mathcal T}_r$ of the radial motion \eqref{Taurdef} is given by
\begin{eqnarray}
\label{Taurdef1}
{\mathcal T}_r&=&\frac{\pi}{\zeta_+}\nonumber\\
&=&\frac{ {\mathcal D}(p)\pi}{\omega}\left[1-\frac{p(2p-1)(2p-3)}{2(p-1){\mathcal D}^2(p)}\, e^2 +O(e^4)\right]\,,\nonumber
\end{eqnarray}
so that the redshift invariant turns out to be
\beq
z_1=\frac{{\mathcal T}_r}{T_r}
=\frac{\sqrt{2}{\mathcal D}(p)}{\omega(3-2p)}\left[1+\frac{2p}{{\mathcal D}^2(p)}\, e^2 +O(e^4)\right] \,.
\eeq


\subsection{Type II orbits}

For type II orbits we have instead
\begin{eqnarray}
\label{phidichitypeII}
t(\chi) &=&\frac{1}{\sqrt{2}}\left[\chi-\sqrt{\frac{1+e}{1-e}} {\rm arctan} (\psi_1)\right]
\,,\nonumber\\
\phi(\chi) &=&-\frac{\chi}{2}\,.
\end{eqnarray}
Thus during one period of the radial oscillation, $\phi$ decreases by $\pi$, requiring two revolutions in $\chi$ for the $x$-ellipse to undergo one full revolution.
The behavior of $\phi$ as a function of $\chi$ is shown in Fig.~\ref{fig:phi_vs_chi} (b) for a selected value of $e$.

The period of radial motion is given by 
\beq
T_r=\sqrt{2}\pi\left(1-\frac12\sqrt{\frac{1+e}{1-e}}\right)\,,
\eeq
while the azimuthal one is $\Phi=-\pi$, implying that $\Omega_r=-2\Omega_\phi$ and $K=-\frac12$.
Finally, 
\beq
{\mathcal T}_r=\frac{\pi}{\omega}\sqrt{\frac{1-3e}{1-e}}\,.
\eeq

\subsection{Type III orbits}

Finally, for type III orbits 
\begin{eqnarray}
\label{phidichitypeIII}
t(\chi)&=&\frac{1}{\sqrt{2}}\left[\chi-\frac{\sqrt{(p-1)^2-e^2}-p}{\sqrt{1-e^2}} {\rm arctan} (\psi_1)\right]
\,,\nonumber\\
\phi(\chi) &=&-\frac{\chi}{2}+{\rm arctan} (\psi_2)\,.
\end{eqnarray}
The behavior of $\phi$ as a function of $\chi$ is shown in Fig.~\ref{fig:phi_vs_chi} (c) for selected values of $p$ and $e$.

The period of radial motion is given by 
\beq
T_r=\sqrt{2}\pi\left[1-\frac{p-\sqrt{(p-1)^2-e^2}}{2\sqrt{1-e^2}}\right]\,,
\eeq
while the azimuthal one is $\Phi=0$, implying that $\Omega_\phi=0$ and $K=-1$.
Finally, 
\beq
{\mathcal T}_r=\frac{\pi}{\zeta_-}\,.
\eeq

\section{Gyroscope precession along Newtonian-like elliptic orbits}

An adapted frame along type I orbits (those we are mainly interested in, because they contain circular orbits in the zero eccentricity limit) is obtained by boosting the frame adapted to the static observers
\begin{eqnarray}
\label{boostedframe}
e(U)_a &=& e(u)_a +\frac{U+u}{1-U\cdot u}(U\cdot e(u)_a)\nonumber\\
&=& e(u)_a+\gamma(U,u)\nu(U,u)_a \, u \nonumber\\
&& +\frac{\gamma(U,u)^2}{\gamma(U,u)+1}\nu(U,u)\nu(U,u)_a
\end{eqnarray}
where $U$ is decomposed in its components along $u$ and orthogonally to it
\begin{eqnarray}
U &=&\gamma(U,u)[u+\nu(U,u)]\equiv \gamma [u +\nu^ce(u)_c]\nonumber\\
&=& \gamma [u +\nu \hat \nu^ce(u)_c]\,,
\end{eqnarray}
with the notation (occasionally abbreviated below by suppressing the relative observer labels $(U,u)$ for simplicity)
\beq
\hat \nu(U,u)=\cos \alpha\, e(u)_1+\sin\alpha\, e(u)_2\,,
\eeq
for the relative velocity unit vector.
Explicitly,
\begin{eqnarray}
\gamma(U,u)&=&\tilde E\,,\\
\nu(U,u)&=&\frac{\omega U_r}{\tilde E\sqrt{2}}e(u)_1 +\frac{\tilde L -2\tilde E \sinh^2 r}{\sqrt{2}\tilde E \sinh r \cosh r}e(u)_2\,,\nonumber
\end{eqnarray}
implying a constant magnitude for $\nu(U,u)$,
\beq
||\nu(U,u)||=\frac{\sqrt{\tilde E^2-1}}{\tilde E}\,,
\eeq
so that,  $\hat \nu(U,u)=\nu(U,u)/||\nu(U,u)||$ and
\begin{eqnarray}
\sin \alpha &=& \frac{\tilde L -2\tilde E \sinh^2 r}{\sqrt{2}\sqrt{\tilde E^2-1} \sinh r \cosh r}\,,\nonumber\\
\cos \alpha &=& \frac{\omega U_r}{\sqrt{\tilde E^2-1}\, \sqrt{2}}\,.
\end{eqnarray}

We find $e(U)_3=e(u)_3$ and
\begin{eqnarray}
e(U)_1&=& \gamma \nu \cos \alpha \, u +[\sin^2\alpha +  \gamma\cos^2 \alpha]\, e(u)_1\nonumber\\
&&+(\gamma-1)\sin\alpha \cos \alpha\, e(u)_2\,, \nonumber\\
e(U)_2&=& \gamma \nu \sin \alpha \, u +  (\gamma-1)\sin \alpha\cos \alpha\, e(u)_1\nonumber\\
&& +[\cos^2\alpha +  \gamma \sin^2\alpha ]\, e(u)_2 \,,
\end{eqnarray}
with both the above expressions of the type $e(U)_a=\Gamma_a ({\mathcal N}_a u+\hat {\mathcal N}_a )$, $a=(1,2)$.
This frame turns out to be a Frenet-Serret frame (degenerate, since the curvature $\kappa$ of these geodesic orbits is zero), with the single nonvanishing torsion
\beq
\label{tau1sol}
\tau_1(r)=- \frac{\omega [2\cosh^4r-2(\tilde E+ \tilde L+1)\cosh^2r+2 \tilde E+\tilde L]}{2\sinh^2r\cosh^2r}  \,,
\eeq
and the following evolution equations along $U$ 
\begin{eqnarray}
\frac{D e(U)_1}{d\tau}&=& \tau_1 e(U)_2\,,\nonumber\\ 
\frac{D e(U)_2}{d\tau}&=& -\tau_1 e(U)_1\,,\nonumber\\
\frac{D e(U)_3}{d\tau}&=& 0\,.
\end{eqnarray}
In the circular case ($\cosh^2r=p$) the torsion \eqref{tau1sol} reduces to $(\tau_1)_{\rm (circ)}=  -\omega$ \cite{Iyer:1993qa}.

Denoting by 
\beq
\label{tau1def}
\tau_1=\frac{d\Psi}{d\tau}
\eeq
the rate of gyroscope precession as a function of the proper time, its conversion in $\chi$ is given by
\begin{eqnarray}
\frac{d\Psi}{d\chi}&=&\tau_1\frac{d\tau}{d\chi}=\frac{\omega\sqrt{1-e^2}}{2\zeta}\left[
\frac{\tilde L+2 \tilde E}{2p}\right. \nonumber\\
&& \left. -\frac12\frac{\tilde L}{1-p+e\cos\chi}-\frac{1}{1+e\cos\chi}
\right]\,,
\end{eqnarray}
so that the total spin precession angle for type I orbits accumulated over a radial period is
\beq
\label{Psidef}
\Psi= 2 \int_0^\pi \frac{d\Psi}{d\chi} d\chi=
-\frac{\omega\pi}{\zeta}
=-\omega{\mathcal T}_r
\,,
\eeq
which is negative like the azimuthal accumulated angle \eqref{periods}.

However, this precession incorporates the rotation of the cylindrical polar coordinate frame in the azimuthal direction, so to remove this, corresponding to comparing the spin direction with the Cartesian-like frame of Eq.~\eqref{exey}, one must subtract $\Phi$ from $\Psi$. The ratio between that net precession angle and the net change in the orbital angle is
\beq
\label{psidef}
\psi=1-\frac{\Psi}{\Phi}=1-\frac{\omega}{2\zeta}\,,
\eeq
The net average precession angle per azimuthal revolution is then  $\Delta \phi_{\rm (prec)}=-2\pi\psi$,
which in the circular limit reduces to
\beq
\Delta \phi_{\rm (prec,circ)}=-2\pi \left[1-\frac12  {\mathcal D}(p)\right]\,,
\eeq
in agreement with \cite{Iyer:1993qa}. 

\begin{figure}
\includegraphics[scale=0.35]{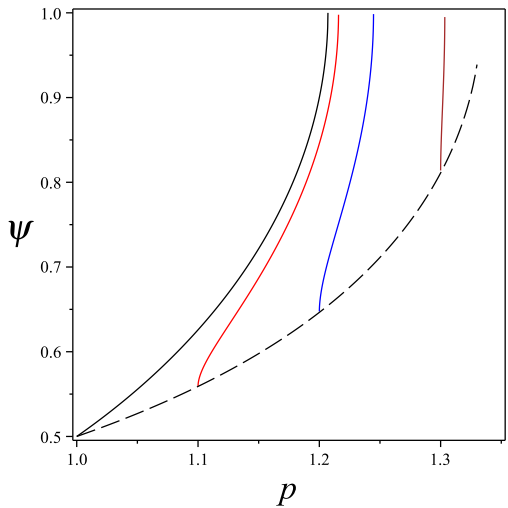}
\caption{The behavior of the spin precession invariant \eqref{psidef} as a function of $p$ for type I orbits is shown for fixed values of $e=[0,0.1,0.2,0.3]$, ordered from left to right.
The minimum value of $\psi$ for every fixed value of $e$ (dashed curve) is obtained by evaluating Eq.~\eqref{psidef} at the limiting value $p=1+e$ which yields $\psi_{\rm min}=1-\frac12\sqrt{{(4-3p)}/{(2-p)}}$.
}
\label{fig:5}
\end{figure}

The behavior of the spin precession invariant as a function of $p$ for selected values of $e$ is shown in Fig.~\ref{fig:5}. 
Since $\psi$ is positive, the spin advances in the direction of the rotation of the orbit as in the black hole case. In the limit of zero radius $(e,p)=(0,1)$, one has $\psi=1/2$ which means that after one revolution the planar projection of the spin advances in the azimuthal direction of motion by 180 degrees with respect to the coordinate grid.


\subsection{Parallel transported frame}

Clearly, the frame $e(U)_a$ is only one of the possible frames which can be adapted to the geodesic world lines by boosting the frame adapted to static observers. One can still spatially rotate this frame by an arbitrary angle and then identify, for example, a parallel propagated frame along geodesics with a suitable choice of that angle. We have recently solved this problem in Ref.~\cite{Bini:2018zfu}.

Rotating the boosted frame \eqref{boostedframe} around $e(U)_3$ by an angle $\beta(r)$ using the radial coordinate as a parameter along the geodesic leads to a parallel transported frame $\{E(U)_a\}$
\begin{eqnarray}
E(U)_1&=&\cos\beta\, e(U)_1+\sin\beta\, e(U)_2
\,,\nonumber\\
E(U)_2&=&-\sin\beta\, e(U)_1+\cos\beta\, e(U)_2
\,,\nonumber\\
E(U)_3&=&e(U)_3
\,.
\end{eqnarray}
For noncircular orbits where $r$ can be used to parametrize the orbit, one finds the parallel transport angle, using Eq.~\eqref{tau1def}, satisfies 
\beq
\beta(r)=-\int^r\frac{\tau_1(r)}{U^r}dr\,,
\eeq
whose explicit expression is given in Eq.~(B11) of Ref.~\cite{Bini:2018zfu}.

\section{Concluding remarks}

We have studied the precession of a test gyroscope moving on a Newtonian-like elliptic geodesic on a $z=$ constant hypersurface in a G\"odel spacetime. The analysis has been conveniently carried out by using an eccentricity ($e$) and semi-latus rectum ($p$) parametrization of the orbits, which is new in this case and is closely related to the analogous parametrization used in the case of bound equatorial motion around a  black hole. 
For the precession angular velocity as well as all the various related quantities (coordinate-time  and proper-time period of the radial motion, frequency of the azimuthal motion, the full variation of the azimuthal angle over a period, etc.) we have provided analytical expressions as well as power series expansions in the (small) eccentricity, in order to read our results as small corrections to circular motion showing what kind of modifications arise off the circular case. In fact, the latter case has been widely studied in the literature, in contrast from the elliptic-like orbits which has never been carefully explored before.
We have shown that for these elliptic-like orbits there is no periastron advance, unlike the black hole case. 
Furthermore, we have discussed the geometric properties of a parallel propagated frame along these geodesics, recently found in a previous paper~\cite{Bini:2018zfu}.
Finally, we expect that this work will be relevant for the study of perturbations to this spacetime induced by a particle moving along an elliptic-like geodesic. Computing the backreaction on the metric within the gravitational self-force scheme will be a challenge for future work.

\appendix

\section{Kundt's form of the G\"odel metric}

The form \eqref{met_godel} of the metric is related to the original metric in Cartesian-like coordinates $(T,X,Y,Z)$ adapted to the Bianchi type VIII homogeneity subgroup
\beq
\label{met_godel_b}
ds^2=-dT^2+dX^2-\frac12 e^{2\sqrt{2}\omega X}dY^2+dZ^2-2e^{\sqrt{2}\omega X}dTdY\,,
\eeq
where the map to cylindrical-like coordinates $(t,r,\phi,z)$ is given by~\cite{Hawell}
\begin{eqnarray}
e^{\sqrt{2}\omega X}&=&\cosh 2r +\cos \phi \sinh 2r
\,,\nonumber\\
\omega Y e^{\sqrt{2}\omega X}&=&\sin \phi \sinh 2r
\,, \nonumber\\
\tan \frac12\left(\phi+\omega T +\sqrt{2}t \right)&=&e^{-2r} \tan \frac{\phi}{2}
\,,\nonumber\\
Z&=&\frac{\sqrt{2}}{\omega}z\,.
\end{eqnarray}
Kundt \cite{kundt} introduced the following coordinate transformation (see also Ref. \cite{pfarr})
\beq
T=t\,,\quad
\sqrt{2}\omega X=-\ln(\sqrt{2}\omega y)\,,\quad
Y=\sqrt{2}x\,,\quad
Z=z\,,
\eeq
which brings the metric \eqref{met_godel_b} in the simplest form 
\beq
\label{met_godel_k}
ds^2=-\left(dt+\frac1{\omega y}dx\right)^2+\frac1{2\omega^2y^2}(dx^2+dy^2)+dz^2\,.
\eeq
The geodesic $4$-velocity vector then reads
\begin{eqnarray}
U&=&-\frac{\sqrt{2}}{C}\left(\frac{y'}{2}-y\right)\partial_t
+\frac{\sqrt{2}}{C}\omega y(y'-y)\,\partial_x\nonumber\\
&+&\frac{\sqrt{2}}{C}\omega y(x-x')\,\partial_y
+\frac{d}{C}\,\partial_z\,,
\end{eqnarray}
with normalization condition ($U^\alpha U_\alpha=-\mu^2$)
\beq
(x-x')^2+(y-y')^2=\frac12(y')^2-d^2-C^2\mu^2\,,
\eeq
where $x'$, $y'$, $C$ and $d$ are integration constants.
Therefore, timelike geodesics ($\mu^2=1$) on a hyperplane $z=$ constant ($d=0$) are all circular at radius 
\beq
\label{rdef}
r=\sqrt{\frac12(y')^2-C^2}\,.
\eeq

\subsection{Gyroscope precession}

To study the precession of a test gyroscope along planar geodesics in the metric \eqref{met_godel_k} it is convenient to introduce polar-like coordinates 
\beq
x = x'+r\cos\phi\,,\qquad
y = y'+r\sin\phi\,,
\eeq
with $r$ given by Eq. \eqref{rdef}.

We start with an orthonormal frame adapted to the static observers with 4-velocity 
$u=e_0=\partial_t$
\begin{eqnarray}
\label{thdgodel}
e_1&=&-\sqrt{2}\cos\phi\,\partial_t+\sqrt{2}\omega(y'+r\sin\phi)\partial_r
\,,\nonumber\\
e_2&=&-\sqrt{2}\sin\phi\,\partial_t-\frac{\sqrt{2}\omega}{r}(y'+r\sin\phi)\partial_\phi
\,,\nonumber\\
e_3&=&\partial_z
\,,
\end{eqnarray}
with respect to which the orthonormal components of $U$ are
\beq
U=\frac{y'}{\sqrt{2}C}\,e_0-\frac{r}{C}e_2\,.
\eeq
Boosting the frame \eqref{thdgodel} into $LRS_U$ leads to
\beq
\label{boosted_frame_godel}
E_a=e_a+\frac{U+e_0}{1-U\cdot e_0}(U\cdot e_a)\,,
\eeq
where $E_1=e_1$ and $E_3=e_3$ are invariant. 
This frame is a (degenerate) Frenet-Serret frame with only nonvanishing torsion $\tau_1=-\omega$.
Therefore, the total spin precession angle is given by $\Psi=-\omega\tau$, in agreement with Eq. \eqref{Psidef} for type I orbits.

\section{Visualizing precession effects}

Consider the Newtonian-like parametrization of a precessing ellipse in polar coordinates $(r,\phi)$ in a plane for one oscillation of the radial variable, i.e.,
\beq
 r=r(\chi)=\frac{p}{1+e\cos\chi}\,,\qquad \phi=\phi(\chi)\,, \qquad \chi\in[0,2\pi]\,,
\eeq
where $0< e<1$.
The instantaneous ellipse corresponding to a point on this orbit at the parameter value $\chi$ which is precessing (rotating around the origin) to generate the orbit is defined by
\beq
\label{fictelli}
r=r(\phi)=\frac{p}{1+e\cos(\phi-(\phi(\chi)-\chi ))}\,, \qquad \phi\in[0,2\pi]\,.
\eeq
The point on this ellipse for which $\phi=\phi(\chi)$ has $r=r(\chi)$ and so coincides with the actual orbit point for this value of $\chi$. The periastron (minimum radius) of the instantaneous ellipse lies at $\phi=\phi_{\rm(prec)}(\chi)= \phi(\chi)-\chi$ so the rate of its precession and thus of the entire ellipse is 
$\phi_{\rm(prec)}(\chi)/d\chi=d\phi(\chi)/d\chi-1$. 

\begin{figure}
\includegraphics[scale=0.7]{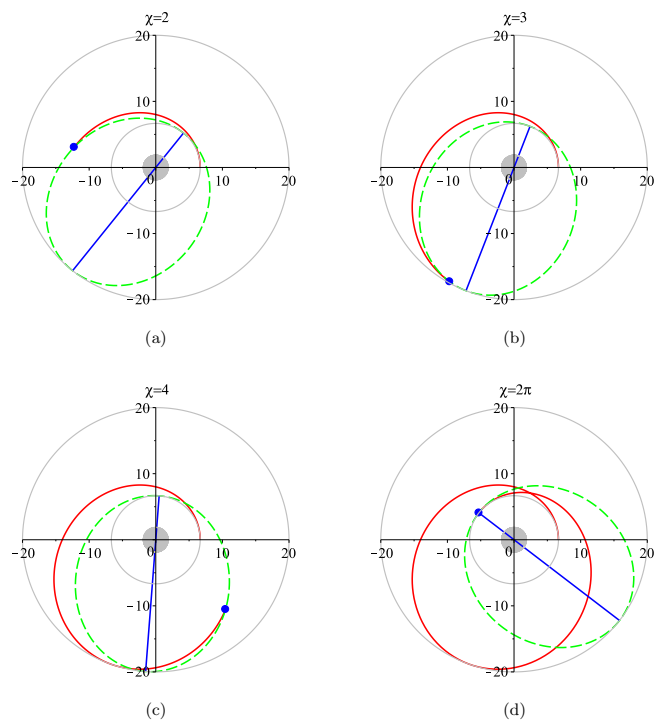}
\caption{
The ``instantaneous ellipse"  (dashed loop) for a bound equatorial orbit in a Kerr spacetime with parameters $a/M=0.5$, $p=10$, $e=0.5$.
}
\label{fig:kerr}
\end{figure}

This representation of bound orbits (or unbound for $e\ge1$) can be used for any classical central force in the plane; for an inverse square central force no precession occurs leading to exact conic section orbits.
This is particularly useful in visualizing how general relativity affects the conic section orbits of Newtonian gravity, as modeled by the equatorial plane timelike geodesics of the Kerr spacetime, where one may apply this in the usual Boyer-Lindquist spherical-like coordinates $(t,r,\theta,\phi)$ at $\theta=\pi/2$. As one approaches the black hole, the conic section orbits precess faster and faster in the same sense as the orbital motion. Showing the symmetry axis through the periastron and focus of the instantaneous conic section in an animation of an orbit together with its instantaneous conic section helps to visualize better its orientation and precession.
Fig.~\ref{fig:kerr} illustrates the case for a bound equatorial plane orbit in the Kerr spacetime~\cite{Bini:2016iym}, clearly showing the advancing periastron of the counterclockwise motion of the orbit.


\begin{figure}
\includegraphics[scale=0.8]{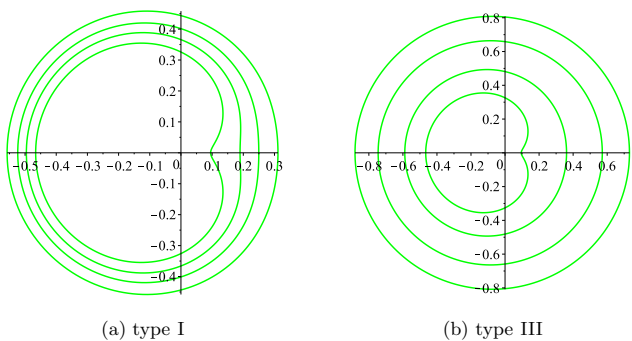}
\caption{
The shape of the instantaneous $x$-ellipse is shown for $e=0.1$ and selected values of $p$ for orbits of type I (panel (a), $p=[1.11,1.14,1.17,1.21]$) and III (panel (b), $p=[1.11,1.25,1.5,1.8]$).
}
\label{fig:fict_ellipses}
\end{figure}

The orbital equation in Newtonian gravity for $r=r(\phi)$ is equivalent to a quadratic potential well problem in $u=1/r$ with sinusoidal solutions which leads immediately to conic section solutions for $r$. In the G\"odel case instead the radial equation of motion for $x(\tau)=\cosh^2(r(\tau))$ is equivalent to a quadratic potential well problem with sinusoidal solutions, suggesting the application of the precessing ellipse representation of the orbit in the $(x,\phi)$-coordinate grid as perhaps being useful.
This leads to the parametrization  Eq.~\eqref{rdichirel}.
The instantaneous $x$-ellipse corresponding to a point on the orbit at the parameter value $\chi$ which is precessing (rotating around the origin) to generate the orbit is defined by
\beq
\label{fictelli2}
r=r(\phi)
={\rm arccosh}\sqrt{\frac{p}{1+e\cos(\phi-(\phi(\chi)-\chi ))}},
\eeq
with $\phi\in[0,2\pi]$ and $\phi(\chi)$ given by Eqs.~\eqref{phidichitypeI}, \eqref{phidichitypeII} and \eqref{phidichitypeIII} for the various orbit types.
Obviously this transformation from $x$ to $r$ deforms the ellipse into a quite different looking deformed conic loop in the $(r,\phi)$ coordinate plane, even cardioid-like in shape, with a concave inward indentation surrounding the periastron.
The shape is always cardioid-like for type II orbits where $p=1+e$, for every fixed value of $e$.  For type I and III orbits 
the shape is also cardioid-like (like the orbits of Fig.~\ref{fig:2}) for all points near the boundary $p=1+e$ which are approximately below the line $p=1+1.5 e$, and elliptic-like above this line.
The deformation of the $x$-ellipse orbit for a representative value of $e$ and various values of $p$ is shown in Fig.~\ref{fig:fict_ellipses}.


\begin{figure*}
\includegraphics[scale=1]{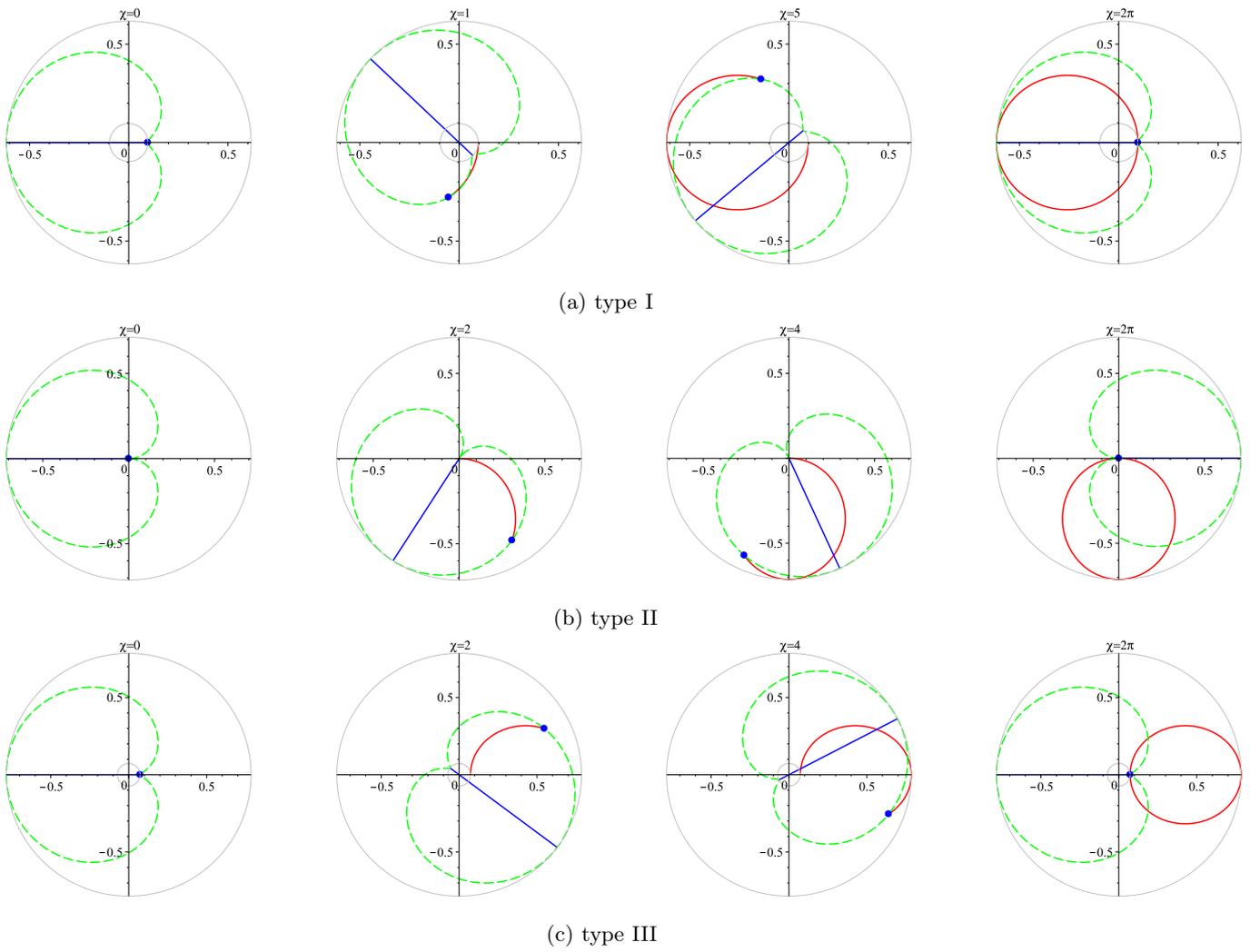}
\caption{
The instantaneous $x$-ellipse (dashed loop) is shown for orbits of each type at selected values of $\chi$.
The parameters are chosen as in Fig.~\ref{fig:2}, but the initial conditions are taken at the periastron ($\chi$=0) instead of the apoastron, so that $\phi(\chi=0)=0$ as in Fig.~\ref{fig:phi_vs_chi}.
}
\label{fig:wobble}
\end{figure*}

The situation is illustrated in Fig.~\ref{fig:wobble} for the various orbit types for a single radial oscillation.
For type I orbits, the instantaneous $x$-ellipse wobbles around the true orbit, returning then to the initial position with no net precession. For type II orbits it rotates counterclockwise for a change in $\phi$ by $\pi$, requiring a second revolution in $\chi$ (and negative values of $r$ corresponding to the reflection across the vertical axis) to smoothly retrace the next half revolution in $\pi$. 
Finally for type III the $x$-ellipse precesses a full loop in the direction of motion during one loop of the actual orbit.


\end{document}